\newcount\eqnno \eqnno=0 
\def\numeqn{\global\advance\eqnno by 1 \eqno(\the\eqnno)} 

\def\etal{{\it et al. }}
\def\nbar{\bar N}

\documentstyle{mn}
\input psfig

\title[Galton--Watson and gravitational clustering]{Galton--Watson branching processes and the growth of gravitational clustering}

\author[Ravi K. Sheth]
       {Ravi K. Sheth \\ 
        Berkeley Astronomy Deptartment, Univeristy of California, 
        Berkeley, CA 94720}
      
\begin{document}

\maketitle

\begin{abstract}
The Press--Schechter description of gravitational clustering from an initially 
Poisson distribution is shown to be equivalent to the well studied 
Galton--Watson branching process.  This correspondence is used to provide a 
detailed description of the evolution of hierarchical clustering, including a 
complete description of the merger history tree.  The relation to branching 
process epidemic models means that the Press--Schechter description can also be 
understood using the formalism developed in the study of queues.  The queueing 
theory formalism, also, is used to provide a complete description of the merger 
history of any given Press--Schechter clump.  In particular, an analytic 
expression for the merger history of any given Poisson Press--Schechter clump 
is obtained.  This expression allows one to calculate the partition function of 
merger history trees.  It obeys an interesting scaling relation; the partition 
function for a given pair of initial and final epochs is the same as that for 
certain other pairs of initial and final epochs.  

The distribution function of counts in randomly placed cells, as a function of 
time, is also obtained using the branching process and queueing theory 
descriptions.  Thus, the Press--Schechter description of the gravitational 
evolution of clustering from an initially Poisson distribution is now complete. 
All these interrelations show why the Press--Schechter approach works well in a 
statistical sense, but cannot provide a detailed description of the dynamics of 
the clustering particles themselves.  One way to extend these results to more 
general Gaussian initial conditions is discussed. 

\end{abstract} 
\begin{keywords} galaxies:  clustering -- galaxies:  evolution -- galaxies:  
formation -- cosmology:  theory -- dark matter.
\end{keywords}
\maketitle

\section{Introduction}
This paper is mainly concerned with providing a complete description of  
gravitational clustering from an initially Poisson distribution.  This is not 
because one thinks it likely that the initial conditions for gravitational 
clustering in our universe were Poisson.  Indeed, measurements of the spectrum 
of temperature fluctuations in the microwave background suggest otherwise.  At 
present, however, analytic understanding of the growth of clustering from more 
general Gaussian initial conditions is not as detailed as for the Poisson case 
studied here.  Thus, the Poisson model serves as a convenient toy model with 
which to study the evolution of nonlinear clustering.  Many of the results 
obtained in this paper should provide at least qualitative insight into the 
evolution of clustering from more general initial conditions. 

The Press--Schechter approach allows one to estimate the distribution of the 
masses of virialized clumps at a given epoch directly from the initial density 
distribution.  It is based on the hypothesis that, provided the initial 
velocities are sufficiently small (i.e., that the initial field is sufficiently 
cold), overdense regions in the initial density field will eventually collapse 
to form nonlinear structures.  Thus, by studying the statistics of overdense 
regions in the initial field, one can infer properties of the nonlinear 
distribution (Press \& Schechter 1974; Bond \etal 1991; Lacey \& Cole 1993).  
One of the main simplifying assumptions of the Press--Schechter approach is 
that the overdense regions are assumed to collapse spherically.  While this may 
be a good approximation in the mean, $N$-body simulations of gravitational 
clustering from initially scale-free Gaussian random fields show that the 
collapse of overdense regions is seldom exactly spherical. Nevertheless, the 
Press--Schechter mass functions for these initially scale-free Gaussian fields 
have been shown to be in good agreement with the distribution of clump sizes 
that are measured in relevant $N$-body simulations (Efstathiou \etal 1988; 
Lacey \& Cole 1994).  

The Press--Schechter excursion set description of clustering from an initially 
Poisson distribution of identical particles has been derived (Epstein 1983; 
Sheth 1995).  The probability that a Poisson Press--Schechter clump has $N$ 
particles is 
\begin{equation}
\eta(N,b) = {(Nb)^{N-1}{\rm e}^{-Nb}\over N!}, 
\label{borel}
\end{equation}
where $N\ge 1$ and $0\le b<1$, and $b$ is related to the Press--Schechter 
overdensity threshold $\delta_{\rm c}$:  
\begin{equation}
b = 1/(1+\delta_{\rm c}) .
\label{bdelta}
\end{equation}
For an initially cold Poisson distribution, the threshold overdensity 
$\delta_{\rm c}$ decreases as the Universe expands in such a way that $b=0$ 
initially, and $b\to 1$ as the clustering develops.  If clumps collapse 
spherically in a universe with critical density, and both growing and decaying 
modes are present in the linear perturbation theory, then 
$\delta_{\rm c} = (5/3)\,1.69/a$, where $a$ is the expansion factor  
(e.g. Bond \etal 1991; Lacey \& Cole 1993).  
Thus, $b$ changes rapidly at first, but at late times its evolution is slowed 
by the expansion of the Universe.  (See section~2.3 in Sheth~1995 for another 
description of the evolution of $b$ that shows this same behaviour.)   
Since $b$ is known to increase monotonically with time, it will be treated as 
a psuedo-time variable in the remainder of this paper.    

Equation~(\ref{borel}) is known as a Borel distribution (Borel 1942).  Epstein 
(1983) derived this distribution by studying the properties of level excursions 
of a Poisson distribution.  His approach is the discrete analog of that which 
was used later by Bond \etal (1991) in their analysis of the initially Gaussian 
random fields.  The Borel distribution can also be derived using simple 
`cloud--in--cloud' conditional probabilities for a Poisson distribution 
(Sheth 1995).  This approach is the discrete analog of Jedamzik's (1995) 
treatment of the Gaussian case.      

If the probability that a randomly chosen clump contains $N$ particles is a 
Borel distribution, then the probability that a randomly chosen particle is in 
such a clump is $N\eta(N,b)/\langle N\rangle = (1-b)N\eta(N,b)$, since the 
average number of particles in a Borel clump is $\langle N\rangle = 1/(1-b)$.  
In the limit of large $N$ and small $\delta_{\rm c}$, Stirling's approximation 
for the factorial term implies that 
\begin{eqnarray}
(1-b)\, N\, \eta(N,b) &=& 
{\delta_{\rm c}\over (1+\delta_{\rm c})}\,
\left({N\over 1+\delta_{\rm c}}\right)^{N-1} \,
{{\rm e}^{-N/(1+\delta_{\rm c})}\over (N-1)!} \nonumber \\
&\to& {\delta_{\rm c}\over\sqrt{2\pi N}} \ \exp 
\left(-{N\,\delta_{\rm c}^2\over 2}\right) 
\label{limn0}
\end{eqnarray}
(Epstein 1983; Sheth 1995).  The final expression is precisely that which 
obtains for a Gaussian density field with white noise initial fluctuations 
(e.g. Bond \etal 1991).  This shows that the Poisson distribution studied in 
the remainder of this paper can be thought of as the discrete analog of the 
white noise Gaussian studied by Bond \etal (1991), and by 
Lacey \& Cole (1993, 1994).

Sections 2 and 3 study other derivations of the Borel distribution.  These 
derivations provide new insight into the accuracy and applicability of the 
Press--Schechter approach.  An analytic expression that is, essentially, the 
partition function that describes all possible merger histories of a given 
Press--Schechter clump is derived in Section 3.  Its properties are consistent 
with those inferred previously (Sheth 1995).  In particular, it is consistent 
with the physical requirement that, in the limit of very small time steps, the 
probability that a clump has two progenitors should be an infinitesimal of 
smaller order than the probability that it has three, or more, progenitors.  
Section 3 also shows that the partition function, and so the growth of 
clustering, satisfies an interesting scaling relation.  In Section 4, this 
scaling relation is exploited to provide insight into the details of the growth 
of hierarchical clustering.  Section 4 also shows that this Poisson 
Galton--Watson, Press--Schechter description is in qualitative agreement with 
the results of the numerical algorithm developed by Kauffmann \& White (1993).  
Section 5 discusses ways in which the branching process extension of the 
Press--Schechter approach that is developed in this paper can be extended to 
provide a detailed description of clustering from initially Gaussian random 
fields.  Appendices A, B and C provide details of some of the calculations.  

Appendix D contains a derivation of the distribution of counts in randomly 
placed cells by extending some of the ideas developed in this paper.  Although 
all the results of this paper are independent of those derived in Appendix D, 
it has been included because the final result in it (equation~\ref{ppsd}) is 
known to be accurate.  Thus, Appendix D provides one natural way in which the 
usual Press--Schechter analysis may be extended to provide additional 
information about the evolution of nonlinear gravitational clustering.  

\section{A relation between the spread of disease and the Press--Schechter 
approach}
Consider a disease that is spread in accordance with the following model.  
Assume that, initially, there is a single carrier of the disease.  Assume that 
this carrier is capable of infecting others, and that the probability that it 
infects $k$ others is given by a Poisson distribution.  If the initial carrier 
is thought of as belonging to the zeroth generation, then each of these newly 
infected carriers belongs to the first generation.  Assume that each of the 
members of the first generation is, in turn, capable of infecting still others, 
who will make up the second generation.  The members of the second generation 
infect still others who make up the third generation, who infect still others, 
and so on.  Assume that, in any generation, the probability that a carrier is 
able to infect $k$ others is given by a Poisson distribution that is specified 
by the parameter, $b$, say.  It is possible that, by chance, none of the 
carriers in the $n$th generation infects any new members.  In this case the 
number of members in the $(n+1)$th generation is zero.  If this should happen, 
the spread of the disease will be said to be halted, and we can ask for the 
probability that there were $N$ people infected in total, including the initial 
member in the zeroth generation.  

This model for the spread of disease has been studied in some detail.  It is a 
Galton--Watson branching process in which the distribution of the number of 
progeny of a given member of each generation is a Poisson distribution with 
mean $b<1$ (see, e.g., Harris 1963).  This Galton--Watson process has an 
analytic solution.  The probability that $N$ people were infected in total is 
given by a Borel distribution with parameter $b$ (Otter 1949; Good 1960; 
Consul 1989).  

It is possible to use this Poisson Galton--Watson branching process to study 
the growth of gravitational clustering from an initially Poisson distribution.  
Consider a randomly chosen point in a Poisson distribution; this point 
comprises the zeroth generation.  All points that are within a given 
`contagious' volume, say $v_{\rm c}$, around this point will be considered to 
be infected.  Since the distribution is Poisson, the probability that this 
point is able to infect $k$ others is Poisson, and the parameter that specifies 
this Poisson distribution is related to the distance from the carrier out to 
which the disease is contagious.  For convenience, assume that $v_{\rm c}$ is 
defined so that the parameter of the Poisson distribution is 
$b=\bar nv_{\rm c}$ where $\bar n$ is the the average density.  These $k$ 
points make up the members of the first generation.  However, each member of 
the first generation will also have been able to `infect', say, $j$ others, 
corresponding to the $j$ neighbours that could have been (Poisson distributed) 
within $v_{\rm c}$ from it.  The set of all these neighbours of all the $k$ 
members of the first generation makes up the second generation, and so on.  It 
is clear that this situation is similar to the one described in the previous 
paragraph.  This means that this Galton--Watson model for clustering from an 
initially Poisson distribution implies that the distribution of nonlinear clump 
sizes is a Borel distribution.  In other words, this Galton--Watson model 
provides the same description of nonlinear clustering as the better known 
Press--Schechter type analyses described earlier.  

\begin{figure}
\centering
\mbox{\psfig{figure=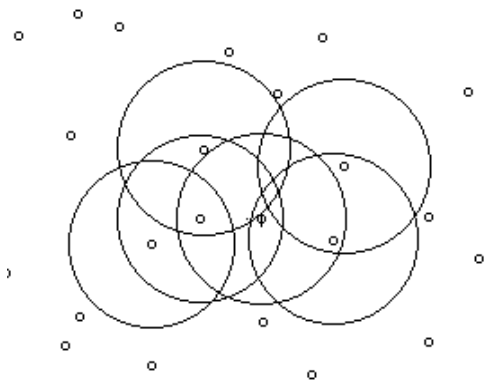,height=5cm}}
\caption{ 
A clump with six member particles identified in the initial particle 
distribution using the friends-of-friends percolation model.  In the 
percolation model, clumps merge with each other as the link length used to 
define friends-of-friends increases.  The overlapping of volumes makes the 
percolation model difficult to treat analytically.}
\label{perco}
\end{figure}

For this description to be exactly like the Galton--Watson process described 
above, we must assume that the probability that one of the members in the first 
generation has $j$ neighbours within $v_{\rm c}$ from it is independent of the 
fact that it is one of $k$ particles within $v_{\rm c}$ from the initial 
particle in the zeroth generation.  This is the same as assuming that the 
volume $v_{\rm c}$ centered on the initial particle does not even partially 
overlap the volume $v_{\rm c}$ centered on each of the $k$ members in the first 
generation.  Clearly, this assumption is false; the assumption that the volumes 
never overlap is a gross simplification.  Nevertheless, we will continue 
considering this simplified model, since it provides the same mass function as 
the Press--Schechter Borel distribution.  

One can argue that the simplification which enabled us to pose the problem in 
terms of this Galton--Watson model suggests one way in which the 
Press--Schechter mass functions could be modified.  Accounting for the fact 
that it is possible for the volume $v_{\rm c}$ centered on the initial particle 
to partially overlap the volume $v_{\rm c}$ centered on one of the $k$ members 
in the first generation, and so on, is the problem known as 
friends--of--friends percolation.  Clearly, percolation is distinct from the 
Galton--Watson process described above.  The percolation model as formulated 
here can be developed as an alternative model for the growth of clustering.  In 
principle, friends--of--friends clump mass functions can be calculated from the 
initial distribution; they are functions of the percolation link length (which 
is simply related to the size of the `contagious' volume $v_{\rm c}$), and are 
different from the clump mass functions determined using the Galton--Watson 
model.  Moreover, mergers are also well-defined in the percolation model, 
as is the partition function of merger trees.  Thus, the percolation model is
at least as well defined as the Press--Schechter excursion set model.  However, 
at present, neither the distribution of friends--of--friends clump sizes, nor 
the associated merger probabilities can be calculated analytically.  Since the 
Galton--Watson model provides the same distribution of clump masses as do the 
excursion set, or the cloud-in-cloud, analyses of the Press--Schechter 
approach, the percolation model will not be considered further in this paper.

\begin{figure}
\centering
\mbox{\psfig{figure=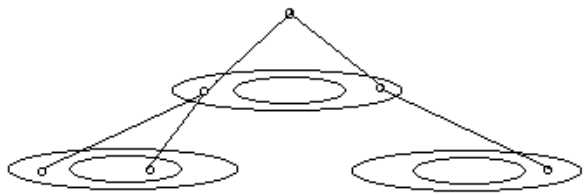,height=3.5cm}}
\caption{ 
A branching process model of hierarchical clustering.  The number of particles 
in each oval is a random variable determined by the (Poisson) initial 
conditions.  The two ovals for each particle represent two different epochs, 
corresponding to two different link lengths, or two different over-density 
thresholds, with oval size increasing (over-density threshold decreasing) with 
time.  In this example, the clump of six particles was composed of four single 
particles and one pair at the earlier epoch.  The fact that different volumes 
are assumed to not overlap makes this branching process analytically tractable.}
\label{gwbp}
\end{figure}

\section{The Galton--Watson description of hierarchical clustering}
The excursion set formulation of the Press--Schechter theory shows clearly how 
to formulate and solve for a description of merging and hierarchical clustering 
(Bond \etal 1991; Lacey \& Cole 1993).  Essentially, the merging problem 
reduces to solving a two-barrier problem that is analogous to the one-barrier 
level crossing problem that was solved to obtain the Press--Schechter 
multiplicity function.  For an initially Poisson distribution the two-barrier 
problem has also been solved.  The probability that, at the epoch $b_1$, one of 
the progenitors of a clump with  $k$ particles at the epoch $b_2>b_1$, was of 
size $j\le k$, is easily related to the conditional distribution 
\begin{eqnarray}
f(j,b_1 \vert k,b_2) &=& {k\choose j}\,{j^j\over k^k}\,
\left({b_1\over b_2}\right)^{j-1}k\left({b_2-b_1\over b_2}\right)\nonumber \\ 
&&\quad\times\ 
\left[k\left({b_2-b_1\over b_2}\right) + (k-j){b_1\over b_2}\right]^{k-j-1},
\label{fjk}
\end{eqnarray}
where $b_1 < b_2$ and $k\ge j$ (equation~40 in Sheth 1995).  Here, 
$f(j,b_1 \vert k,b_2)$ is the probability that a particle which is in a 
$k$-particle clump at the epoch $b_2$ was in a $j$-particle clump at the epoch 
$b_1$.  The statements $f(j,b_1 \vert k,b_2)$ are independent of what happened 
at times previous to $b_1$ and also of what will happen at times later than 
$b_2$.  Appendix B shows that, in the appropriate limits (i.e., $k\gg 1$, 
$j\gg 1$, and $k-j\gg 1$, and also $b_1\to 1$ and $b_2\to 1$), 
equation~(\ref{fjk}) reduces to equation~(2.15) of Lacey \& Cole (1993) 
(also see Fig.~2 and associated algebra in Sheth 1995).

Following Lacey \& Cole (1993), equation~(\ref{fjk}) can be manipulated to 
provide an expression for the probability that a clump with $k$ particles at 
the epoch $b_2$ was formed from other subclumps that merged with a subclump 
that had exactly $j$ particles at the epoch $b_1$.  However, it does not 
provide any information about the distribution of sizes of the other 
subclumps, other than the restriction that the total number of particles in 
those other subclumps must sum to $k-j$.  Another way of thinking of 
$f(j,b_1\vert k,b_2)$ is to note that it is obtained without consideration of 
the different ways in which the object of size $j$ (at the epoch $b_1$) could 
have merged with other objects to form the final object of size $k$ (at the 
epoch $b_2$).  If the merging process is visualized as a tree (so trees having 
different numbers of branches, branch sizes and branching points describe 
different merger histories), a useful quantity is the probability that a 
particular tree structure, rather than any other, occurs.  So, the problem is 
to solve for what is, essentially, the partition function for various tree 
structures.  

Recall that the statements $f(j,b_1 \vert k,b_2)$ are independent of 
what happened at times previous to $b_1$ and also of what will happen at 
times later than $b_2$.  So, to solve for the entire merger history tree 
(i.e., for many $b_1 < b_2 < \cdots$) one need only know how to solve 
for the tree structure at any given two epochs, say, $b_1$ and 
$b_2$.  Therefore, in what follows the later epoch, $b_2$, is 
referred to as the final epoch.  Notice also that the statements 
$f(j,b_1 \vert k,b_2)$ imply that it is possible to calculate the 
partition function for a tree with $k$ particles without explicitly considering 
the partition function for trees with $l\ne k$ particles.  So, for any given 
pair of epochs $b_1$ and $b_2$, and for any tree whose final size (i.e., at the 
epoch $b_2$) is $k$, the problem is to calculate all possible tree 
configurations, and to assign to each configuration the probability that it 
actually occurs.  The elements of the set of all possible $k$-particle tree 
configurations are simply the various ways of partitioning the integer $k$ 
(physically, this is what is required by mass conservation).   

Let $p(1^{n_1}2^{n_2}\cdots\,k^{n_k} | k)$, with $n_1+\cdots +n_k = m$ and 
$\sum_{j=1}^k j\,n_j = k$, denote the probability that the final clump with 
$k$ particles at $b_2$ had $m$ progenitors at $b_1$, of which $n_1$ were 
singles, $n_2$ were pairs, $n_j$ were subclumps with $j$ particles each, and so 
on.  Note that $p(1^{n_1}2^{n_2}\cdots\,k^{n_k} | k)$ is independent of the 
order of $\{1^{n_1}2^{n_2}\cdots\,k^{n_k}\}$ since different permutations of 
the `branches' should all have the same probability of occurring.  The problem 
is to calculate this probability for given $k, m, b_2, b_1$ and 
$\{n_1,n_2,\cdots,n_k\}$.  To date, it has not proven possible to calculate 
this probability exactly from the excursion set description itself 
(Kauffmann \& White 1993; Sheth 1995).  In this section we will use the 
Galton--Watson interpretation of the clustering process to obtain probabilities 
for the various partitions of $k$.  

The Galton--Watson process can be used to formulate a description of the merger 
history of a clump as follows.  As before, consider the single member of the 
zeroth generation.  This person has a number of children, and they each have 
some number of children, and so on down the family tree.  If the probability a 
given member of the tree has $n$ children is given by a Poisson distribution 
with parameter $b_2$, then we have the same branching process as before.  The 
probability that such a family tree has $N$ members in total, after which the 
family died out, is given by a Borel distribution with parameter $b_2$.  Now 
consider the more complicated case in which some of the children are male and 
some female.  Assume that the probability that a given member on the tree has 
$n_1$ daughters is a Poisson distribution with parameter $b_1$ and the 
probability of having $n_2$ sons is a Poisson distribution with parameter 
$b_2-b_1$.  This is the same as assuming that the probability that a given 
member on the tree has $n$ children is given by a Poisson process with 
parameter $b_2$, and the probability that $n_1$ are daughters and $n_2=n-n_1$ 
are sons is given by a Binomial distribution where the probability that any 
given child is female is $p=b_1/b_2$.  Consider the family tree of such a 
process, and assume that the initial ancestor was male.  For such a family 
tree, we can ask for the probability that there were $N_1$ females and 
$N_2=N-N_1$ males in total in the family tree.  This branching process is a 
special case of that studied by Good (1960).  

Now imagine drawing the family tree.  Use dots to represent children of either 
sex, but only draw lines between parents and their daughters.  Then the tree 
will consist of a number of groups, in which each member of the group is 
connected to other members of the same group, but not to members from different 
groups.  Call these groups subfamilies.  (The tree is inherently sexist, as the 
first member at the head of every new subfamily is always male.)  

We will characterize trees by the distribution of the sizes of the subfamilies 
in them.  Let $n_i$ denote the number of subfamilies each with $i$ members.  
Then we can ask for the probability $p(1^{n_1}\cdots\,k^{n_k}|k)$ that a tree 
with $k$ members has exactly $m$ subfamilies (so $n_1+\cdots +n_k = m$), and 
that there were $n_1$ singles, $n_2$ pairs, $n_j$ $j$-tuples, and so on.  
Clearly, this characterization of a family tree (by the size and number of 
subfamilies in it) is exactly analagous to the distribution of subclump sizes 
for a given Press--Schechter clump.  It provides us with probabilities for the 
various partitions of $k$.  In particular, this extension of the Poisson 
Galton--Watson branching process requires that 
\begin{eqnarray}
p(1^{n_1}2^{n_2}\cdots\,k^{n_k}|k) \!&=&\! 
{\eta(1,b_1)^{n_1}\eta(2,b_1)^{n_2}\cdots\,\eta(k,b_1)^{n_k} \over 
\eta(k,b_2)}\ \nonumber \\
\!&&\!\quad\times\quad {m!\over n_1!n_2!\cdots\,n_k!}\quad\times \nonumber \\
\!&&\!\quad {1\over m}\,{\left[k(b_2-b_1)\right]^{m-1}\,
{\rm e}^{-k(b_2-b_1)}\over (m-1)!} ,
\label{prtfnc}
\end{eqnarray}
where $\eta(l,b)$ is the Borel distribution with parameter $b$ 
(eq.~\ref{borel}), and $n_j$ denotes the number of subfamilies in the tree that 
have exactly $j$ members (with the usual convention that $0!$ = 1).  Since this 
branching process is related to the gravitational clustering process, the 
probability that a clump of size $k$ at the epoch $b_2$ had the $m$ progenitors 
$n_1,n_2,\cdots ,n_k$ at the epoch $b_1$ is given by equation~(\ref{prtfnc}).  
Thus, equation~(\ref{prtfnc}) can be used to generate the partition function of 
merger history trees.  It is the main result of this paper.

Appendix B shows that this partition function is consistent with the merger 
probabilities of equation~(\ref{fjk}).  It also provides additional insight 
into the origin of the various terms in equation~(\ref{prtfnc}).  Appendix C 
describes a queueing process that provides another way to derive this 
expression.  

\subsection{Some properties of the partition structure}
Some limits of equation~(\ref{prtfnc}) are worth studying.  As $b_2\to b_1$, 
$p(1^{n_1}\cdots\,k^{n_k}|k)\to 0$, except when $m=1$, for which 
$p(k^1|k)\to 1$ as required.  As $b_1\to 0$, then $\eta(l,b_1)\to 0$ except 
for $l=1$, so that 
\begin{eqnarray}
\lim_{b_1\to 0} p(1^{n_1}\cdots\,k^{n_k}|k) &=& p(1^k|k) \nonumber \\
&=& {(kb_2)^{k-1}{\rm e}^{-kb_2}\over k!}{1\over \eta(k,b_2)} = 1 .
\end{eqnarray}
In other words, in the limit as $b_1\to 0$, all progenitors are certainly 
single particles, which is also the expected result.  

The probability $n(m|k)$ that a clump of size $k$ at the epoch $b_2$ has $m$ 
progenitors at the epoch $b_1$ is given by summing 
$p(1^{n_1}\cdots\,k^{n_k}|k)$ over all distinct sets of $m$ integers that add 
up to $k$.  Now, any set specified by $\{n_1,\cdots\,,n_k\}$ with 
$n_1+\cdots +n_k=m$ can be written explicitly in terms of its $m$ members as 
$\{l_1,\cdots\,l_m\}$.  Thus, $l_1+\cdots\,+l_m = \sum_{j=1}^k jn_j = k$.  So, 
\begin{eqnarray} 
n(m|k) &=& \sum_{m\ {\rm parts.}}
{\eta(1,b_1)^{n_1}\cdots\,\eta(k,b_1)^{n_k} \over \eta(k,b_2)}\ 
{m!\over n_1!\cdots\,n_k!}\nonumber \\
&&\qquad\qquad\times\quad {1\over m}\,{\left[k(b_2-b_1)\right]^{m-1}\,
{\rm e}^{-k(b_2-b_1)}\over (m-1)!} \nonumber \\
&=& \sum_{m\ {\rm parts.}} 
\left({1^{1-1}\over 1!}\right)^{n_1}\cdots\,
\left({k^{k-1}\over k!}\right)^{n_k} \nonumber \\
&&\qquad\qquad\times\quad{b_1^{1n_1+\cdots\,+kn_k-n_1-\,\cdots\,-n_k}\over 
{\rm e}^{\,1n_1b_1+\cdots\,+kn_kb_1}} \nonumber \\
&&\qquad\qquad\times\quad {k!\over (kb_2)^{k-1}\,{\rm e}^{-kb_2}}\ \times\ 
{m!\over n_1!\cdots\,n_k!} \nonumber \\
&&\qquad\qquad\times\quad {1\over m}\,{\left[k(b_2-b_1)\right]^{m-1}\,
{\rm e}^{-k(b_2-b_1)}\over (m-1)!} \nonumber \\
&=& {k!\over m!}{1\over k^{k-m}}\left({b_1\over b_2}\right)^{k-m}
\left({1 - {b_1\over b_2}}\right)^{m-1} \nonumber \\
& &\qquad\qquad \times\ \sum_{{\rm dist. perms.}} {l_1^{l_1-1}l_2^{l_2-1}\cdots\, l_m^{l_m-1}\over l_1!l_2!\cdots\,l_m!} 
\nonumber \\
&=&{k!\over m!}\,{1\over k^{k-m}}\left({b_1\over b_2}\right)^{k-m}
\left({1 - {b_1\over b_2}}\right)^{m-1} {m\over k}{k^{k-m}\over (k-m)!} 
\nonumber \\
&=&{k-1\choose m-1}\left({b_1\over b_2}\right)^{k-1-(m-1)}
\left(1-{b_1\over b_2}\right)^{m-1}.
\label{nmk}
\end{eqnarray}
The sum in the third equality is over all sets of $m$ integers which satisfy 
$l_1+\cdots\,+l_m=k$, and over all the distinct permutations of each set 
(which accounts for the multinomial factor $m!/n_1!\cdots\,n_k!$).  The second 
from last equality follows from a combinatorial identity (note the similarity 
to the Borel--Tanner distribution of equation~\ref{btanner}; also see 
Sheth \& Saslaw 1994).  The final expression for $n(m|k)$ is the same as that 
obtained previously (eq.~52 in Sheth 1995).  

It is easy to see that this expression is sensible.  Referring back to the 
Galton--Watson family tree description, recall that the head of the tree is 
always the initial male, and other males start subfamilies within the tree.  
So, the number, $m$, of subfamilies within a family tree having $k$ members in 
total is equal to the number of male members in the tree.  Now, the probability 
any given child is male is $q=1-p=(b_2-b_1)/b_2$.  Since it is certain that one member of the tree is a male (the head of the family tree is always a male), 
the probability that there are $m$ males in a family with $k$ members should be 
$[(k-1)!/(k-m)!(m-1)!]\,p^{k-m}\,q^{m-1}$.  This is identical to the Binomial 
distribution of equation~(\ref{nmk}).  Alternatively, this expression for the 
number of subclumps of a $k$-sized clump, which is the same as the number of 
sons in a $k$ sized family, can be obtained directly from Good's (1960) 
treatment of the Galton--Watson process with many types of progeny.  The brute 
force calculation is given in Appendix A.  It, too, shows that $n(m|k)$ is 
given by equation~(\ref{nmk}).

Equation~(\ref{nmk}) is also consistent with the physical requirement that, in 
the limit of very small time steps, the probability that a clump has two 
progenitors should be an infinitesimal, the probability that the clump has 
three progenitors should be an infinitesimal of the next higher order, and so 
on.  It also implies that massive clumps form preferentially from massive 
progenitor clumps:  on average, the size, at some earlier epoch, of the largest 
progenitor clump of a clump at some given final epoch depends significantly on 
how massive the final clump is relative to the characteristic mass at the final 
epoch (eq.~55 in Sheth 1995). 

In the limit of large $k$ and large subclumps $l_i$, equation~(\ref{prtfnc}) 
has an interesting interpretation.  Stirling's formula for the factorials 
implies that 
\begin{eqnarray}
p(l_1,\cdots\ ,l_m|k) &=& 
{l_1^{l_1-1}\cdots\,l_m^{l_m-1}\over l_1!\cdots\,l_m!}\,
{k!\over k^{k-1}} \,{k^{m-1}\over n_1!\cdots\,n_k!} \nonumber \\
&&\qquad\times\quad \left({b_1\over b_2}\right)^{k-m} 
\left(1-{b_1\over b_2}\right)^{m-1} \nonumber \\
&\to&{k^{m-1}\over n_1!\cdots\,n_k!}\,
\left({b_1\over b_2}\right)^{k-m}\!\!\left(1-{b_1\over b_2}\right)^{m-1} 
\nonumber \\
&&\qquad \times\quad {\sqrt{2\pi} k^{3\over 2}\over (2\pi)^{{m\over 2}}}\,
\prod_{i=1}^m {1\over l_i^{3/2}} \nonumber \\
&\to& {1\over n_1!\cdots\,n_k!}\,\left({b_1\over b_2}\right)^{k-m} 
\!\!\left(1-{b_1\over b_2}\right)^{m-1}\nonumber \\
&&\qquad \times\quad {1\over (2\pi k)^{(m-1)/2}}\,
\prod_{i=1}^m {1\over x_i^{3/2}} ,
\end{eqnarray}
where we have defined $x_i\equiv l_i/k$.  The first term on the right is the 
combinatorial term that is included to insure that each distinct combination of 
subclumps is counted only once.  The second set of terms accounts for the 
probability that there are exactly $m$ subclumps.  The final product term is 
the most interesting.  It shows that, for a given value of the final clump size 
$k$, and if $m\ll k$, the case in which the subclumps are all approximately the 
same size is much less likely than the case in which one of the subclumps is 
very much more massive than all the others.  

The partition function (eq.~\ref{prtfnc}) exhibits an interesting scaling.  It 
depends on the initial and final epochs, $b_1$ and $b_2$, only through the 
combination $p=b_1/b_2$.  As one would expect, this is also true of the simpler 
statement given in equation~(\ref{fjk}).  This means that the probability 
$p(l_1\cdots ,l_m|k)$, given the two epochs $b_1$ and $b_2$, will be the same 
as the probability $p'(l_1\cdots ,l_m|k)$, given $b'_1$ and $b'_2$, provided 
that $b_1/b_2 = b'_1/b'_2 = p$.  In this sense, the clustering evolves in a 
self--similar fashion.  This scaling can be exploited when comparing 
equation~(\ref{prtfnc}) with merger histories of clumps in $N$-body 
simulations.  When $b_1\to 1$ and $b_2\to 1$, the ratio $b_1/b_2 = 
(1+\delta_{{\rm c}2})/(1+\delta_{{\rm c}1})\to 
1 + \delta_{{\rm c}2} - \delta_{{\rm c}1}$.  Thus, the scaling in 
$b_1/b_2$ corresponds to the scaling in $(\delta_{{\rm c}1}-\delta_{{\rm c}2})$ 
discussed, for example, by Bond \etal (1991) and by Lacey \& Cole (1993).  

\section{Applications}
\begin{figure}
\centering
\mbox{\psfig{figure=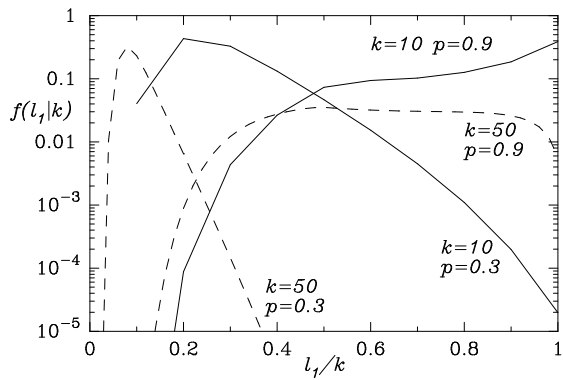,height=5cm,bbllx=90pt,bblly=435pt,bburx=465pt,bbury=706pt}}
\caption{
Probability, $f(l_1|k)$, that the largest subclump of a $k$ particle clump has 
$l_1$ particles, for two choices of the ratio $p=b_1/b_2$ of the initial and 
final epochs, and for two choices of $k$.  Solid lines show the distribution 
for $k=10$ with $p=0.3$ and 0.9.  Dashed lines show the distribution for 
$k=50$, with the same values of $p$. }
\label{lproj}
\end{figure}

One example of the type of statistical question that we can now answer is as 
follows.  Suppose we are interested in the distribution of sizes of the largest 
progenitor clump at some epoch $b_1$, of a given clump at the epoch $b_2$.  
This might be of interest, for example, in studies of the Butcher--Oemler 
effect (Bower 1991; Kauffmann 1995).  It is straightforward to compute the 
relevant sums over the partition function numerically.  As an example, 
Fig.~\ref{lproj} shows the probability that the largest subclump of a clump 
with $k$ particles has $l_1$ particles, for two choices of $k$, and for two 
choices of the ratio $p=b_1/b_2$ of the initial and final epochs.  

Three features of Fig.~\ref{lproj} are obvious.  First, for given $k$, the 
curves depend strongly on $p$, the ratio of the two epochs $b_1$ and $b_2$.  As 
$p$ increases, the curves peak at higher values of $l_1/k$.  This means that 
the largest progenitor of a given clump is a smaller fraction of the total mass 
as the time between the final epoch and the epoch at which the progenitors were 
identified increases.  This simply reflects the fact that the clustering is 
hierarchical; small clumps merge to form big clumps, and on average, clumps 
were smaller in the more distant past than they are at present.  Second, the 
curves depend strongly on $k$, the number of particles in the final clump.  For 
a given value of $p$, curves with higher values of $k$ peak further towards the 
left.  That is, for a given value of $p$, the largest progenitor of a massive 
clump is more likely to be a smaller fraction of the total mass than is the 
largest progenitor of a less massive clump.  In this sense, for any choice of 
initial and final epochs (because the partition function only depends on the 
ratio $b_1/b_2$), more massive clumps always appear to have assembled a larger 
fraction of their mass more recently than less massive clumps.  Lacey \& Cole 
(1993) show that this is also what happens in the Gaussian case.  Furthermore, 
it is consistent with the analytical result that, for a clump that has $k$ 
particles in total, the average number of particles in a progenitor subclump 
(not necessarily the largest subclump) is $k/[p + k(1-p)]$ (Sheth 1995).  

So, we expect the average size, and the most probable size, of the largest 
progenitor clump to decrease as $p$ decreases.  However, based on 
Fig.~\ref{lproj}, the scaling property of the partition function, and the fact 
that $b$ evolves quickly initially and slower at later times (see discussion 
following eq.~\ref{bdelta}), we can also study the rate with which this size 
decreases.  As a specific example, consider two clumps, each of size of $k$.  
Assume that one of the clumps is completely assembled at epoch $b_2$ and the 
other at $b'_2< b_2$ (the primed clump was assembled earlier than the unprimed 
clump).  Now consider the progenitors of the unprimed clump that are identified 
at the epoch $b_1=pb_2$.  The partition function (eq.~\ref{prtfnc}) specifies 
this distribution of progenitor subclumps.  So, we can compute, e.g., the 
average size of the largest progenitor subclump.  Now consider the primed 
clump.  The scaling property of the partition function shows that its subclump 
distribution will be the same as that of the unprimed clump at the epoch when 
$b'_1 = pb'_2<b_1$, where $p=b_1/b_2$ has the same value as for the unprimed 
clump.  Consider the average size of the largest progenitor of these two clumps 
as a function of `lookback time' from the epochs $b_2$ and $b'_2$.  

Since $p$ is the same for the two clumps, they will have the same average size 
for the largest subclump at the epochs $b_1$ and $b'_1<b_1$.  However, the 
lookback time for the unprimed and the primed clumps is the time corresponding 
to $b_{21}=b_2-b_1$ and $b'_{21}=b'_2-b'_1$, respectively.  Since $b$ changes 
ever more slowly as it increases, $b'_{21}<b_{21}$.  This implies that the 
average size of the largest progenitor of the primed clump (which was assembled 
earlier) decreases more rapidly than it does for the unprimed clump (which was 
assembled later).  In other words, when phrased in terms of lookback time, the 
evolutionary history of a clump of size $k$ depends on when it was first 
assembled.  

As discussed above, the evolutionary histories of clumps, when phrased in terms 
of lookback time, depend on the rate of change of $b$.  In the previous 
paragraph we were able to draw conclusions about the dependence of evolution on 
formation epoch because the rate of change of $b$ is a function of epoch.  
However, as noted in the introduction, the rate of change of $b$ also depends 
on the background cosmology.  Thus, one also expects the evolutionary histories 
of clumps to be sensitive to the background cosmology.  For example, the $b(t)$ 
curves for different cosmologies show that a clump of a given mass will have 
formed at a greater lookback time in a low density universe than in one with 
critical density.  These trends, the more recent assembly of larger relative to 
smaller clumps, and the sensitivity of merger histories to the background 
cosmology, have been noted by Lacey \& Cole (1993) and by Kauffmann (1995) in 
their study of clustering from initially Gaussian fields.  The implication that 
similar mass clumps which were assembled at different times have different 
lookback time histories is in qualitative agreement with the numerical, 
Monte--Carlo model used by Kauffmann (1995).  

The third feature that is evident in Fig.~\ref{lproj} is simply that the curves 
are extremely skew.  This means that the average number of particles in the 
largest progenitor subclump is not necessarily a good indicator of the most 
probable number of particles in the largest progenitor.  Therefore, the curves 
of average merger histories given in Fig.~1 of Kauffmann (1995) should be 
treated carefully.  Note that equation~(\ref{prtfnc}) provides an efficient way 
of evaluating the difference between the mean and the most probable sizes.  

In addition to allowing one to calculate the dispersion around the mean history 
of any given clump, the partition function can also be used to compare this 
Galton--Watson Poisson Press--Schechter model with $N$-body simulations.  It is 
also useful to compare the merger histories described by 
equation~(\ref{prtfnc}) with the ad hoc, Monte--Carlo merger histories 
generated by Kauffmann \& White (1993).  This will provide a test of the 
Galton--Watson model developed here, and may also provide some insight into the 
reason for the accuracy of the Kauffmann--White algorithm.  

\begin{figure}
\centering
\mbox{\psfig{figure=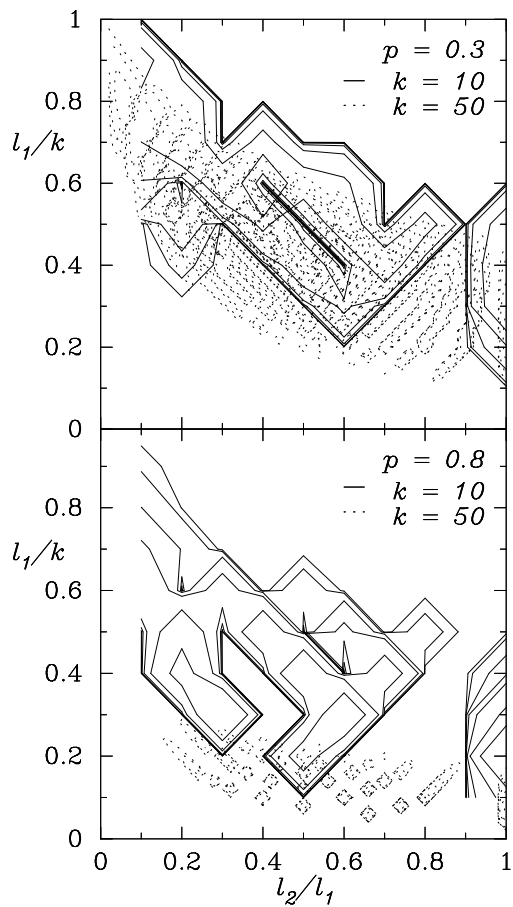,height=12cm,bbllx=86pt,bblly=130pt,bburx=367pt,bbury=631pt}}
\caption{Joint probability that the largest two progenitor subclumps are 
$l_1$ and $l_2$, given that the final clump is of size $k$ as determined by 
the partition function (eq.~5) for $p=b_1/b_2 = 0.8$ (top panel) 
and $p=0.3$ (bottom panel), when $k=10$ (solid contours) and $k=50$ (dotted 
contours).  The results are plotted in terms of the ratio $l_1/k$ versus 
$l_2/l_1$, for ease of comparison with the earlier work described in the text.  
The contours are at probability levels of $0.0001$, $0.0003$, $0.001$, $0.003$, 
$0.01$, $0.03$, and $0.1$. }
\label{pninj}
\end{figure}

To effect this comparison, we will plot the ratio of the largest progenitor 
clump to the total mass $l_1/k$, versus the ratio of the second largest 
progenitor to the largest progenitor, $l_2/l_1$.  Figures~2 and 3 in 
Kauffmann \& White (1993) show examples of such plots.  Fig.~\ref{pninj} 
(of this paper) shows what is essentially the joint probability that the 
largest two progenitor subclumps are $l_1$ and $l_2$, given that the final 
clump is of size $k$, as determined by the partition function 
(eq.~\ref{prtfnc}) for $p=b_1/b_2 = 0.8$ (top panel) and 0.3 (bottom panel), 
and for $k=10$ (solid contours) and $k=50$ (dotted contours).  Large values of 
$p$ correspond to large time differences between the initial and final epochs.  
The results are plotted in terms of the ratio $l_1/k$ versus $l_2/l_1$, for 
ease of comparion with the work of Kauffmann \& White (1993).  The contours 
are at 0.0001, 0.0003, 0.001, 0.003, 0.01, 0.03, and 0.1.  The figure shows 
that, for a given value of $k$, most of the clumps lie along a relatively 
narrow band in the $(l_1/k),(l_2/l_1)$ plane.  The location of the band 
depends on the values of $k$ and $p$.  

Fig.~\ref{pninj} is qualitatively similar to Figs.~2 and~3 in Kauffmann 
\& White (1993).  At small lookback times (large $p$, top panel) the band lies 
at somewhat larger values of $l_1/k$ than at larger lookback times (small $p$, 
bottom panel).  This simply reflects the fact that, at smaller lookback times, 
a larger fraction of the clump survives relatively intact.  However, the band 
for the clump with larger $k$ is at a lower value of $l_1/k$ than the less 
massive clump.  This is consistent with the faster assembly required by larger 
clumps that we deduced in Fig.~\ref{lproj}.  The top panel also shows that, in 
this (relatively small lookback time) regime, the joint probability 
distribution has one peak at large values of $l_1/k$ (and correspondingly small 
values of $l_2/l_1$), and a broader, not so high peak around the values of 
$l_1\sim k/2$ and $l_2\sim l_1/2$.  Thus, at small lookback times, it appears 
that most clumps grow because a large clump accretes many smaller ones.  Those 
that do not, grow because of the mergers of objects that are approximately the 
same size.  

At larger lookback times (bottom panel), the band is shifted towards lower 
values of $l_1/k$.  Furthermore, the joint probability distribution becomes 
peaked towards the lower right hand corner of the plot, at values of 
$l_2/l_1\approx 1$.  This, too, is sensible, since at large lookback times, 
most clumps have split up into a large number of small, approximately equally 
massive progenitors.  These features are in qualitative agreement with those 
measured in $N$-body simulations (Kauffmann \& White 1993, and references 
therein).

\section{Discussion}
The partition function that describes the relative probabilities of all 
possible merger histories of those Press--Schechter clumps which form as an 
initially Poisson distribution evolves gravitationally, can be written in 
closed form (eq.~\ref{prtfnc}).  The partition function is a function of the 
initial and final epochs (denoted by $b_1$ and $b_2$, respectively).  Since the 
time evolution of $b$ can be computed (cf. the Introduction) the temporal 
evolution of the partition function is known.  The counts in cells distribution 
associated with this Poisson Press--Schechter distribution can also be computed 
by extending the branching process analogy (Appendix D).  It, too, is a 
function of $b$, so its temporal evolution is also known.  Thus, the 
Press--Schechter description of clustering from an initially Poisson 
distribution is now complete.  

It may be worth pointing out that Appendix B shows that the Poisson 
Galton--Watson branching process specifies the Borel clump size distribution 
(equation~\ref{borel}) and the merger probabilities (equation~\ref{fjk}) 
uniquely, and also allows one to describe the merger tree completely 
(equation~\ref{prtfnc}).  On the other hand, if equations~(\ref{borel}) 
and~(\ref{fjk}) are treated as the only known constraints on the form of the 
merger tree, then the branching process considered in this paper is not the 
only way to construct the merger history trees consistent with these 
constraints (e.g., Sheth 1995).  For instance, one can always construct an ad 
hoc Monte-Carlo scheme, like that of Kauffmann \& White (1993), which satisfies 
equations~(\ref{borel}) and~(\ref{fjk}), and is not necessarily consistent with 
the branching process description of equation~(\ref{prtfnc}).  For this reason 
we have argued that, rather than being completely ad hoc, the branching process 
is, indeed, a reasonable model for the growth of clustering (Figs.~\ref{perco} 
and~\ref{gwbp} and associated discussion).  

A Poisson distribution is similar to a white noise Gaussian random field.  So, 
this result can be related to the evolution of clustering from an initially 
Gaussian density field that has a scale free power spectrum with slope $n=0$.  
Can it be extended to describe the growth of gravitational clustering from 
Gaussian random fields with arbitrary initial power spectra?  

One way in which to do this is as follows.  The Borel distribution reduces to 
the Press--Schechter multiplicity function in the limit as $N\gg 1$ and as 
$b\to 1$.  Similarly, $f(j|k)$ (eq.~\ref{fjk}) reduces to the $n=0$ Gaussian 
result provided that $k\gg 1$, $j\gg 1$, $k-j\gg 1$, and $b_1$ and $b_2\to 1$ 
(Sheth 1995).  For a given density threshold $\delta_{\rm c}$ [eq.~\ref{bdelta} 
shows that $b=1/(1+\delta_{\rm c})$], Gaussian fields with different power 
spectra yield different excursion set Press--Schechter mass multiplicity 
functions.  However, these differences arise solely because the relation 
between the variance and the mass depends on the power spectrum.  When written 
directly in terms of the variance, rather than the mass, the Press--Schechter 
multiplicity functions and the merger probabilities have a universal form that 
is independent of the underlying power spectrum (e.g. Lacey \& Cole 1993).  
This suggests writing the partition function of equation~(\ref{prtfnc}) in 
terms of the variance of the Poisson ($n=0$) distribution.  Then, in the large 
$l_i$ and $b\to 1$ limits, equation~(\ref{prtfnc}) should reduce to the 
Gaussian result.  The appropriate Jacobian can then transform the partition 
function (written in terms of the variance) into an expression for the masses 
of the subclumps.  

There may be a more sophisticated way to describe the merger histories of 
clumps that form from Gaussian random fields.  In this paper, branching 
processes and queueing theory were both used to formulate and solve a problem 
that was originally posed in terms of trees associated with random walk 
excursions on a discrete Poisson distribution.  So the question is, Is it 
possible to extend any of these relations to the continuous Gaussian case?  For 
example, is it possible to formulate the excursion sets of random walks 
associated with Gaussian fields, described by Bond \etal (1991) in their 
derivation of the Press--Schechter mass functions, in terms of branching 
processes?  Trees associated with excursions of random walks on Gaussian fields 
are the subject of current interest (Neveu \& Pitman 1989; Le Gall 1993).  The 
$b\to 1$ limit is known as the `critical' Galton--Watson process (the $b<1$ 
case considered in this paper is `subcritical').  Many properties of the tree 
associated with this critical process have been calculated (Aldous 1993 and 
references therein).  The application of these ideas to the Press--Schechter 
description of clustering from arbitrary Gaussian initial conditions is in 
progress.  
 
One final way in which to extend the ideas of this paper to arbitrary initial 
conditions is to note that the usual Press--Schechter mass functions provide a 
good, but by no means perfect, description of $N$-body simulations of 
clustering.  Thus, one might reconsider the percolation model discussed in 
Section~2.  First, one would compare the percolation mass functions (obtained 
numerically) for the Poisson case with those of the branching process (the 
Borel distribution, equation~\ref{borel}).  If the percolation mass functions 
provide an acceptable fit to the simulation results, then one could also 
compute (numerically) and test the percolation merger probabilities and merger 
tree.  Of course, it is trivial to apply the percolation model to arbitrary 
initial conditions.  

The discussion at the end of Section~2 shows clearly that this branching 
process extension of the Press--Schechter approach is correct only in a 
strictly statistical sense.  This is because, unlike in the percolation model 
discussed in Section~2, there is no direct correspondence between particles in 
the initial distribution and particles in clumps.  That is, the Galton--Watson 
branching process, like the excursion set interpretation of the 
Press--Schechter mass function, is a purely statistical model for the growth of 
clustering.  To be useful, it relies on the accuracy and applicability of the 
ergodic hypothesis.  It cannot, and should not, be expected to work on a 
particle-by-particle basis. 

\section*{Acknowledgments}
I thank I. M. and R. G. Hook for much needed encouragement, G. Lemson for 
arranging a trip to the island of Schiermonnikoog where my interest in 
branching process descriptions of clustering began, D. Aldous for stimulating 
discussions, J. Pitman for providing valuable insight and entertaining cricket, 
and the referee, C. Lacey, for suggestions that improved the presentation of 
these results.

\appendix

\section{The Poisson Galton--Watson process with two types of progeny}
The Poisson Galton--Watson process with $m$ different types of progeny has been 
studied by Good (1960).  He shows that if the branching process starts with 
$r_i$ individuals of type $i$, for all $1\le i\le m$, then the probability that 
the whole tree contains exactly $n_1$ of type $1$, $n_2$ of type 2, and so on, 
is equal to the coefficient of 
$$
z_1^{n_1-r_1}z_2^{n_2-r_2}\cdots z_m^{n_m-r_m}
$$
in
$$
f_1^{n_1}\cdots f_m^{n_m}\,\left\vert\!\left\vert \delta_\mu^\nu - {z_\mu\over f_\mu}\,{\partial f_\mu\over \partial z_\nu} \right\vert\!\right\vert ,
$$
where $\delta_\mu^\nu$ is the identity matrix and  
$$
f_\mu(\bmath{z}) = \prod_\nu \exp\Bigl(-a_{\mu\nu}(1-z_\nu)\Bigr)
$$
reflects the fact that the probability that an individual of type $\mu$ has a 
child of type $\nu$ is a Poisson distribution with parameter $a_{\mu\nu}$ 
independently of the other individuals.  

For our problem, $m=2$, and subscript 1 is for girls and subscript 2 is for 
boys.  Then $a_{11} = a_{21} = b_1$ and $a_{12} = a_{22} = b_2-b_1$, and the 
problem is to extract the relevant coefficients of 
\begin{eqnarray}
&&\!\!\!\!\!\!\!\!\!\!\!\!
\Biggl\{\exp\Bigl(n_1\bigl[b_1(z_1-1) + (b_2-b_1)(z_2-1)\bigr] \nonumber \\
&& \qquad + \quad n_2\bigl[b_1(z_1-1) + (b_2-b_1)(z_2-1)\bigr] \Bigr)\Biggr\} 
\nonumber \\
&&\!\!\!\!\!\!\!\!\!\!\times\ \Biggl\{\Bigl[1-z_1b_1\Bigr]\,\Bigl[1-z_2(b_2-b_1)\Bigr] - 
z_2b_1\,z_1(b_2-b_1)\Biggr\} .
\label{good2}
\end{eqnarray}
Suppose that we are interested in the family tree that was started by one male 
ancestor, in which there are $N$ members of the tree in total.  Then we can use 
equation~(\ref{good2}) to calculate the probability of having a tree with 
exactly $N$ members.  Let $n$ denote the number of females in the tree.  Then 
set $n_1=n$ and $n_2=N-n$ and let $r_1=0$ and $r_2=1$.  The answer is obtained 
by summing up all the coefficients of terms of order $z_1^n\,z_2^{N-n-1}$.  
This involves straightforward but tedious algebra which is not reproduced here. 
The result is that the probability of having $N$ members in the tree is 
\begin{eqnarray}
P(N)&=& \eta(N,b_2) \nonumber \\
&&\quad\times\quad \sum_{n=0}^{N-1}{N-1\choose n}
\left({b_1\over b_2}\right)^n\left(1-{b_1\over b_2}\right)^{N-1-n} \nonumber \\
&=& \eta(N,b_2),
\end{eqnarray}
where $n$ denotes the number of girls in the tree.  Clearly, this is the same 
result as that given by equation~(\ref{nmk}) in the main text.  

\section{Merger probabilities and the partition function}
Some of the results of this Appendix were suggested by the recent work of 
Pitman (in preparation).  This Appendix shows explicitly that the partition 
function derived in the main text, equation~(\ref{prtfnc}) and the merger 
probabilities of equation~(\ref{fjk}) are consistent with each other.  
Demonstration of this consistency is useful because, in the appropriate limit, 
the merger rates implied by equation~(\ref{fjk}) are in good agreement with 
$N$-body simulations of gravitational clustering (Lacey \& Cole 1994).  For 
completeness, this limit is derived below.  In addition, an expression for the 
mean number of $j$-sized subclumps of $k$-sized clumps is derived directly from 
the partition function.  The argument is generalized, at the end of this 
Appendix, to obtain an expression for the factorial moments of this 
distribution.  

Lacey \& Cole (1993) define a merger rate by taking the limit as 
$\delta_1\to\delta_2$ in $f(k,\delta_2|j,\delta_1)$.  That is, the merger 
rate is 
\begin{eqnarray}
{{\rm d}P(j\to k\vert\delta)\over {\rm d}\delta}\,{\rm d}\delta &\equiv&
\lim_{\delta_1\to\delta_2} {(1-b_2)\,k\,\eta(k,b_2)\ 
f(j,\delta_1|k,\delta_2)\,\over (1-b_1)\,j\,\eta(j,b_1)} \nonumber \\ 
&=& \lim_{\delta_1\to\delta_2} {k\over (k-j)!}\,{\delta_2\over \delta_1}\,
{(\delta_1 - \delta_2)\over (1+\delta_2)^2}\,\nonumber \\
& &\quad\times\quad \left( {k\over 1+\delta_2}-{j\over 
1+\delta_1}\right)^{k-j-1} \nonumber \\
& &\quad\times\quad {\rm e}^{-{k\over 1+\delta_2}+{j\over 1+\delta_1}} 
\nonumber \\
&\to& {k\over \sqrt{2\pi}(k-j)^{3/2}}\,{{\rm d}\delta\over (1+\delta)}\,
{{\rm e}^{\delta(k-j)/(1+\delta)}\over (1+\delta)^{k-j}} \nonumber \\
&\to&\!\!\! {{\rm d} \delta\over\sqrt{2\pi}}\,
\left({k^2\over(k-j)}\right)^{3/2}\,
{{\rm e}^{-\delta^2 (k-j)/2}\over k^2\,(1+\delta)}
\label{mrgrate}
\end{eqnarray}
(Sheth 1995).  The third expression on the right follows from setting 
$\delta_1=\delta_2=\delta$ and $\delta_1 -\delta_d = {\rm d}\delta$, and 
considering the limit when $k\gg 1$, $j\gg 1$ and $k-j\gg 1$.  Stirling's 
approximation for the factorials simplifies the expression considerably, and 
the final expression follows from assuming $\delta\ll 1$.  Except for the 
$(1+\delta)$ term in the denominator, the final expression is the same as the 
expression derived by Lacey \& Cole (1993).  In their notation 
$S_1\propto 1/j, S_2\propto 1/k$, and $\omega=\delta$, since the relevant 
Gaussian corresponding to the Poisson is the white noise case.  So, in the 
limit where Stirling's approximation for the factorials is valid, and when 
$\delta\ll 1$, equation~(\ref{mrgrate}) here reduces to their equation~(2.17).  
In this limit, the merger rate implied by equation~(\ref{fjk}) describes the 
simulations well (Lacey \& Cole 1994).

One other limit of equation~(\ref{fjk}) is also interesting.  When $k\gg j$, 
Stirling's approximation for $k!$ and $(k-j)!$ in equation~(\ref{fjk}) implies 
that  
\begin{eqnarray}
f(j,b_1 \vert k,b_2)\!\!&\to&\!\!\left(1-{b_1\over b_2}\right)
{j^{j-1}\over (j-1)!}\left({b_1\over b_2}\right)^{j-1}
\!\!{\rm e}^{-j(b_1/b_2)}.
\end{eqnarray}
This shows that the probability that a randomly chosen member of a Borel clump 
that has exactly $k$ members at the epoch $b_2$ was in a Borel clump with 
$j\ll k$ particles at the epoch $b_1$ is, to an excellent approximation, given 
by a Borel distribution with parameter $b_1/b_2$.  When $b_2\to 1$, this means 
that the probability that a particle is in a clump with $j-1$ other particles 
is given by a Borel distribution with parameter $b_1$.  Since the limit 
$b_2\to 1$ is equivalent to requiring $\delta_2\to 0$, this is exactly what is 
required by the derivation of $f(j|k)$ from the two barrier (excursion set) 
problem considered in Sheth (1995).  Another application of Stirling's 
approximation (to the $j!$ term), with the limits $b_1\to 1$ and $b_2\to 1$ 
shows that equation~(\ref{fjk}) is similar to the Lacey \& Cole (1993) 
expression for the white noise Gaussian case.  Fig.~2 in Sheth (1995) also 
shows this to be true.  

Before deriving the merger probabilities $f(j|k)$ of equation~(\ref{fjk}) 
directly from the partition function (equation~\ref{prtfnc}), it is useful to 
consider some combinatorial identities.  First, consider the random variable 
$X$, and assume that the distribution of $X$ is Borel with parameter $b$.  Then 
the distribution of $S_m = X_1 + \cdots + X_m$, where the $X_i$ are independent 
random variables, each drawn from a Borel distribution with parameter $b$, is 
given by the Borel--Tanner distribution.  That is, the probability that 
$S_m = k$ (i.e., the sum of $m$ independent Borel variables equals $k$) is 
\begin{equation}
P(b,S_m=k) = {m\over k}\,{(kb)^{k-m}\ {\rm e}^{-kb}\over (k-m)!} ,
\quad{\rm where}\ k\ge m
\label{btanner}
\end{equation}
(Tanner 1953, 1961).  It is easy to see that this expression is sensible, 
since  
\begin{eqnarray}
&&\!\!\!\!\!\!\!\!\!\!\!\!\sum_{j=1}^{k-m+1} 
\eta(j,b)\ P(b,S_{m-1}=k-j) \nonumber \\
&& = \sum_{j=1}^{k-m+1} {(jb)^{j-1}\,{\rm e}^{-jb}\over j!}\,
{m-1\over k-j}\,{\left[(k-j)b\right]^{k-m}\,{\rm e}^{-(k-j)b}\over (k-j-m+1)!} 
\nonumber \\
&& = (m-1)\ {b^{k-m}{\rm e}^{-kb}\over (k-m)!}\nonumber \\
&&\qquad\times\quad \sum_{i=0}^{k-m}{k-m\choose i}\ 
(1+i)^{i-1}\,(k-1-i)^{k-m-i-1} \nonumber \\
&& = P(b,S_m=k) ,
\end{eqnarray}
as it should.  The final expression follows from Abel's generalization of the 
Binomial theorem (see equations~14 and~20 in section 1.5 of Riordan 1979 or 
equation~46 of Sheth 1995).  Iterating this process shows that 
\begin{eqnarray}
P(b,S_m=k) &=& \!\!\!\!\sum_{l_1=1}^{k-m+1} \eta(l_1,b)\ P(b,S_{m-1}=k-l_1) 
\nonumber \\
&=& \!\!\!\!\sum_{l_1=1}^{k-m+1} \eta(l_1,b) \quad \times \nonumber \\
&&\quad\sum_{l_2=1}^{k-l_1-m+2}\!\!\!\!\eta(l_2,b)\ P(b,S_{m-2}=k-l_1-l_2) 
\nonumber \\ 
&=& \!\!\!\!\sum_{{\rm all}\ m\ {\rm parts.}}\!\!\!\! 
\eta(l_1,b)\eta(l_2,b)\cdots\,\eta(l_m,b) \nonumber \\
&=& \!\!\!\!\sum_{{\rm all}\ m\ {\rm parts.}} \!\!\!\!\eta(1,b)^{n_1} 
\cdots\,\eta(k,b)^{n_k},
\label{pbsmk}
\end{eqnarray}
where $n_1+\cdots+n_k=m$, and $l_1+\cdots+l_m = \sum_{j=1}^k j\,n_j = k$, and 
the sum in the two final expressions is over all distinct ordered partitions of 
$k$ that have exactly $m$ parts.  Notice that not all the terms in this sum are 
different.  For example, when $m=3$ and $k=6$, then the set $\{123\}$ occurs 
six times, and when $m=3$ and $k=7$, then the set $\{223\}$ occurs thrice.  In 
general, a given set $\{n_1,\cdots,n_k\}$, will occur $m!/(n_1!\cdots\,n_k!)$ 
times.  

Let $p(n_1\cdots\,n_k|m,k)$ denote the probability that a given set 
$\{n_1,\cdots,n_k\}$ occured, given that there were exactly $m$ terms which 
added up to $k$.  Then $p(n_1\cdots\,n_k|m,k)$ is given by summing up all the 
terms in equation~(\ref{pbsmk}) corresponding to it, and normalizing by 
$P(b,S_m=k)$.  If it is not certain that there were exactly $m$ terms in the 
partition, then we must multiply $p(n_1\cdots\,n_k|m,k)$ by the probability 
that there were exactly $m$ terms that added up to $k$.  Thus, 
$p(n_1\cdots\,n_k|k)$, which is conditioned on $k$ only and not on $m$ as well, 
is given by an expression like   
$$
{\eta(1,b)^{n_1}\eta(2,b)^{n_2}\cdots\,\eta(k,b)^{n_k}\over 
P(b,S_m=k)}\,{m!\over n_1!\cdots\,n_k!}\ \times\ n(m|k) ,
$$
where $n(m|k)$ denotes the probability that $k$ is the sum of exactly $m$ 
integers.  Now, Appendix A showed that the branching process we are considering 
requires $n(m|k)$ to have the Binomial distribution of equation~(\ref{nmk}).  
Thus, the probability that a given set $\{n_1,\cdots,n_k\}$ occurs is 
\begin{eqnarray}
p(n_1\cdots\,n_k|k) &=& {\eta(1,b_1)^{n_1}\cdots\,\eta(k,b_1)^{n_k}\over 
P(b_1,S_m=k)}\,{m!\over n_1!\cdots\,n_k!}\nonumber \\
&&\quad \times\quad {k-1\choose m-1}\left({b_1\over b_2}\right)^{k-m} 
\left(1-{b_1\over b_2}\right)^{m-1} \nonumber \\
&=& {k!\over m!}\,{(b_2-b_1)^{m-1}\over b_2^{k-1}}\,
{{\rm e}^{kb_1}\over k^{k-m}} \nonumber \\ 
&&\quad\times\quad {m!\over n_1!\cdots\,n_k!}\ \prod_{j=1}^k \eta(j,b_1)^{n_j}.
\end{eqnarray}
Simple algebra shows that this final expression is equivalent to 
equation~(\ref{prtfnc}).  This shows explicitly how the partition function is 
related to sums of Borel-distributed random variables.  

We are finally in a position to show that equation~(\ref{fjk}) follows from 
equation~(\ref{prtfnc}).  The probability that a particle which is chosen at 
random from a clump with exactly $k$ particles at the epoch $b_2$ was in a 
subclump with $j$ particles at the epoch $b_1$ is 
$$
\sum_{{\rm all\ part.}} {j\,n_j\over k}\ p(n_1,\cdots,n_k|k) ,
$$
where the sum is over all partitions of $k$.  This sum can be written as
\begin{eqnarray}
\sum_{{\rm all\ part.}} {j\over k}\ \eta(j,b_1)\,
{\eta(1,b_1)^{n_1}\cdots\,\eta(j,b_1)^{n_j-1}\cdots\,\eta(k,b_1)^{n_k}
\over P(b_1,S_m=k)} && \nonumber \\
\qquad\times\quad {(m-1)!\over n_1!\cdots\,(n_j-1)!\cdots\,n_k!}\ m\,n(m|k) , 
\quad && 
\end{eqnarray}
which is the same as 
\begin{equation}
\sum_{m=1}^{k-j+1} {j\over k}\ \eta(j,b_1)\ 
{P(b_1,S_{m-1}=k-j)\over P(b_1,S_m=k)} \ m\ n(m,b_1|k,b_2) ,
\end{equation}
where it is understood that when $m=1$ then 
$P(b_1,S_{m-1}=k-j) = P(b_1,S_0=0) = 1$, and $P(b_1,S_0=k-j)=0$ for $j\ne k$.  
Writing all the terms in the sum explicitly gives
\begin{eqnarray}
\sum_{m=1}^{k-j+1} {j\over k}\ {(jb_1)^{j-1}{\rm e}^{-jb_1}\over j!}\ 
{m-1\over k-j}\,{[(k-j)b_1]^{k-j-m+1}\,{\rm e}^{-(k-j)b_1}\over (k-j-m+1)!} 
&&\!\!\!\! \nonumber \\
\ \times\ {k\over m}\,{(k-m)!\over (kb_1)^{k-m}\,{\rm e}^{-kb_1}}
\ m\ {k-1\choose m-1}\left({b_1\over b_2}\right)^{k-m}
\left(1-{b_1\over b_2}\right)^{m-1} && \!\!\!\!\nonumber \\
= \sum_{m=2}^{k-j+1} {k\choose j}\,{j^j\over k^k}\,
\left({b_1\over b_2}\right)^{j-1} k\left(1-{b_1\over b_2}\right) 
\quad\times\qquad\qquad\qquad  \ \ \ \ && \nonumber \\ 
\!\!\!\!{k-j-1\choose m-2} \left[k\left(1-{b_1\over b_2}\right)\right]^{m-2}
\left[(k-j)\,{b_1\over b_2}\right]^{k-j-m+1}\!\!\!\!.&&\!\!\!\! 
\end{eqnarray}
The Binomial theorem reduces this final expression to equation~(\ref{fjk}).  
This shows explicitly that 
\begin{equation}
f(j,b_1|k,b_2) = \sum_{{\rm all\ part.}} {j\,n_j\over k}\ p(n_1,\cdots,n_k|k)
\label{meanj}
\end{equation}
as expected.  Thus, the merger probabilities of equation~(\ref{fjk}) can be 
derived directly from the partition function (equation~\ref{prtfnc}), which 
means that the merger probabilities and the partition function are mutually 
consistent.

The steps leading to equation~(\ref{meanj}) imply that the mean number of 
subclumps each having exactly $j$ particles that are incorporated in a clump 
with exactly $k$ particles is 
\begin{equation}
\sum_{{\rm all\ part.}} n_j\ p(n_1,\cdots,n_k|k) = {k\over j}\,f(j,b_1|k,b_2).
\end{equation}
This is the same result as that obtained using a different argument 
(equation~45 in Sheth 1995).  However, with the partition function, it is now 
possible to compute the higher order moments of this distribution as well.  For 
completeness the factorial moments are computed below.  These higher order 
moments are useful for estimating the scatter around the mean number of 
$j$-clumps per $k$-clump.  Also, they may be good discriminators between 
different partition functions that yield the same $f(j|k)$ merger 
probabilities.  The factorial moments are 
\begin{eqnarray}
\left\langle\!\!{n_j!\over (n_j-i)!}\!\right\rangle\!\!\!\! &=&\!\!\!\! 
\sum_{m=i+1}^{k-ij+i} \eta(j,b_1)^i\ {P(b_1,S_{m-i} = k-ij)
\over P(b_1,S_m=k)}\nonumber \\
&&\quad\times\quad {m!\over (m-i)!}\ n(m,b_1|k,b_2) \nonumber \\
\!\!\!\!&=&\!\!\!\! {k!\over j!^i\,(k-ij)!}\,{(j^{j-1})^i\over k^{k-1}} 
\left({b_1\over b_2}\right)^{ij-i}\!
\left[k\left(1 -{b_1\over b_2}\right)\right]^i\!\!\!\!\nonumber \\ 
\!\!\!\!&&\!\!\!\!\ \times\ \left[k\left(1-{b_1\over b_2}\right) + 
(k-ij){b_1\over b_2}\right]^{k-ij-1} \!\!.\!\!\!\!
\end{eqnarray}
Since the partition function is known, it is also straightforward to compute 
`cross--correlation' type moments of the form $\langle n_i n_j\rangle$, and the 
associated factorial moments, though we have not done so here.  

\section{Merging, branching, and the theory of queues}
The Borel distribution (eq.~\ref{borel}) also arises in studies of the 
distribution of waiting times in queues (Borel 1942; Tanner 1953; 
Tanner 1961).  The fact that queueing theories and the Galton--Watson branching 
process are closely related (Kendall 1951) will be exploited in this Appendix. 

Consider a counter at which customers are served.  Assume that customers arrive 
at the counter in a Poisson process with parameter unity (an average of one 
arrival per unit time) and that the service time is the same constant, 
$0\le b\le 1$, for each customer.  If the counter is busy when a customer 
arrives, the customer joins the back of a queue.  So, service is on a 
first-come-first-served basis.  In such a system, we can ask for the 
probability that exactly $N$ customers are served before the queue is first 
emptied, given that there was only one customer in the queue initially.  This 
probability is given by the Borel distribution with parameter $b$ (Borel 1942; 
Tanner 1953; Consul 1989).  The relation of this queue to the Galton--Watson 
branching process described above is clear.  Simply view the $N$ customers that 
were served before the queue was first emptied as the descendents, the ones who 
were `infected' by, the initial customer.  

Now modify the queue system as follows.  Assume, as before, that customers 
arrive at a counter in a Poisson process with unit rate, and that the service 
time is the same constant, say $b_2$, for each customer.  However, in this 
case, the customers form two queues in accordance with the following 
prescription.  If a new customer arrives within a time $b_1\le b_2$ of the most 
recent commencement of service, they join the back of a high priority queue, 
$H$.  If they arrive after later than $b_1$ of the most recent commencement of 
service, they join the back of a low priority queue, $L$.  All customers in 
queue $H$ are serviced, in the order in which they arrived, until the queue is 
empty.  When this happens, the first customer in queue $L$ moves into queue $H$ 
and is serviced immediately.  Thus, while customers within each queue are 
serviced on a first-come-first-served basis, it is possible for some customers 
in queue $H$ to receive service before others in queue $L$, even though they 
may have arrived later.  In this sense, the modified system does not operate on 
a strictly first-come-first-served basis.  

For such a system, we can ask for the probability that exactly $k$ customers 
are served before both queues are completely emptied for the first time, given 
that there was only one customer in queue $H$ and none in queue $L$ initially.  
Clearly, the answer to this question is no different than before:  this 
probability must be given by the Borel distribution with parameter $b_2$.  
However, in the time before both queues were emptied for the first time, having 
serviced exactly $k$ customers, queue $H$ may have been emptied a number of 
times (though certainly not more than $k$ times).  Suppose that it was emptied 
$m$ times.  Define a batch of customers as the number, $l$, of customers served 
between two successive empty periods of queue $H$.  Then we can ask for the 
probability, $p(l_1,l_2,\cdots ,l_m|k)$, that queue $H$ was emptied $m$ times, 
and that customers were served in batches of $l_1,l_2,\cdots ,l_m$ (not 
necessarily in that order), given that $l_1+l_2+\cdots +l_m = k$, and that 
there was only one customer in queue $H$ initially.  Then 
$p(l_1,l_2,\cdots ,l_m|k)$ is the same as that defined in the Galton--Watson 
process considered earlier in this paper.  So, for this queue system, it is 
given by equation~(\ref{prtfnc}). 

In the context of this queue system, equation~(\ref{prtfnc}), $\eta(l,b)$ is 
the Borel distribution with parameter $b$ (eq.~\ref{borel}), and $n_j$ denotes 
the number of times exactly $j$ customers passed through queue $H$ before it 
was emptied, given that $k$ passed through it before both $H$ and $L$ were 
emptied.  The first term on the right of equation~(\ref{prtfnc}) accounts for 
the fact that the probability of serving exactly $l_i$ customers between two 
successive empty periods in the $H$ queue is a Borel distribution, and so it 
weights the occurence of each $l_i$ in the sequence of service batches with the 
probability, $\eta(l_i,b_1)$, that $l_i$ occured.  The second term on the right 
accounts for the different permutations of the different service sequences, 
since  the various permutations of $\{l_1,\cdots,l_m\}$ all contribute to the 
same $p(l_1,l_2,\cdots ,l_m|k)$.  The final term looks like a Poisson weighting 
term, and is obtained by an argument that is similar to that used by 
Tanner (1961) in his elegant derivation of the Borel distribution.  

Consider the case in which both queues are empty, having serviced $k$ customers 
since the last time they were both empty.  If the $H$ queue was emptied $m$ 
times during this busy period, then $m-1$ customers must have passed through 
queue $L$ during this time.  So, we can ask for the probability that $m-1$ 
customers arrived at and passed through queue $L$ during this time.  However, 
in Tanner's (1961) language, not all possible arrival patterns are 
`admissible', since the arrivals in queue $L$ must be consistent with the 
pattern of service times in queue $H$.  Namely, during each of the first $m-1$ 
instants at which queue $H$ is emptied, there must be at least one customer in 
queue $L$, but the $m$th time that $H$ is emptied, queue $L$ must also be 
empty.  The problem is to list all possible sequences of arrivals at queue $L$, 
and then to determine the fraction of these that are admissible.  
 
Recall that customers arrive in a Poisson process with unit rate at a queueing 
system that has constant service time $b_2\le 1$ per customer, and the first 
customer joins queue $H$ without passing through queue $L$.  If a customer 
arrives during the first fraction, $b_1/b_2$, of the service time, then they 
join queue $H$, otherwise they join queue $L$.  If $k$ customers were served in 
total, then there were $k$ opportunities for customers to join queue $L$.  Each 
of these $k$ opportunities was of duration $b_2-b_1$.  Since the arrival of 
customers is random, the probability that $m-1$ customers passed through queue 
$L$ during the time in which $k$ customers were served is given by the Poisson 
distribution, $[k(b_2-b_1)]^{m-1}{\rm e}^{-k(b_2-b_1)}/(m-1)!$.  However, the 
probability that the $m-1$ arrivals were in an admissible order introduces an 
additional factor of $1/m$.  This sets the final term in 
equation~(\ref{prtfnc}).  

In terms of the Galton--Watson branching process considered in the main text 
(in which the probability that any parent has $n_1$ daughters is a Poisson 
distribution with parameter $b_1$, and the probability that that parent also 
had $n_2$ sons is a Poisson distribution with parameter $b_2-b_1$), 
equation~(\ref{prtfnc}) lists the probability that a family of size $k$ is made 
of the $m$ subfamilies (each with a male at the head and only females in the 
subsequent generations), having $l_1,l_2,\cdots ,l_m$ members in each.  So, for 
the gravitational clustering process, the probability that a clump of size $k$ 
at the epoch $b_2$ had the $m$ progenitors $l_1,l_2,\cdots ,l_m$ at the epoch 
$b_1$ is given by equation~(\ref{prtfnc}).  As noted in section~3, 
equation~(\ref{prtfnc}) can be used to generate the partition function of 
merger history trees.

This way of deriving the partition function, by generalizing Tanner's (1961) 
argument, has another connection to previous work.  In effect, it is an 
alternative derivation of the excursion set scaling solution derived in 
section~3.2 of Sheth (1995).  This follows because Tanner showed how his 
queueing system could be formulated in terms of an excursion set process.  Here 
we have described a queueing system that is associated with the Poisson 
Galton--Watson branching process.  Comparison of this queue with the excursion 
set process considered in section~3.2 of Sheth (1995) shows that they are 
equivalent.  

\section{Counts-in-cells from the Galton--Watson branching process}
The branching process extension of the Press--Schechter approach is very 
powerful.  In the main text it was used to provide a description of the merger 
history tree.  However, following a calculation suggested by Consul (1989), it 
is also relatively straightforward to use the Galton--Watson branching process 
to provide a derivation of the distribution of counts in randomly placed cells, 
at any epoch characterized by $b$.  Assume that Press--Schechter Borel clumps 
collapse completely to points, so that if the cluster center is included in a 
cell, all associated particles are also.  Now assume that all clumps evolve in 
accordance with the Press--Schechter description developed in the main text.    

The analogy with the Galton--Watson process is as follows.  The number of 
particles in a given cell is the same as the total number of progeny of the 
Galton--Watson process.  However, unlike the Press--Schechter mass function 
considered in the main text, in this case the number of initial ancestors, 
$X_0$, is not unity.  Rather, it is the number of cluster centers, $X_0=m$, 
say, that happen to be in the cell.  So, for a cell containing $m$ cluster 
centers, rather than considering the total number of progeny of one ancestor 
(the Borel distribution with parameter $b$), we need to calculate the total 
number of progeny given that there are $X_0=m$ ancestors in the zeroth 
generation.  This is the same as the Poisson Galton--Watson process, 
conditioned on the number of ancestors being $X_0=m$, rather than unity.  Both 
the branching process (Consul 1989) and the queueing theory (Tanner 1953) 
formulations of this problem show that the probability that there are exactly 
$N$ particles in the cell given that there are exactly $m$ cluster centers in 
the cell is
\begin{equation}
P(N|m) = {m\over N}\,{(Nb)^{N-m}{\rm e}^{-Nb}\over (N-m)!} ,
\end{equation}
which is the Borel--Tanner distribution of Appendix C.  When $m=1$ it reduces 
to the Borel distribution of equation~(\ref{borel}).  

The final key idea is to assume that, since the initial distribution is 
Poisson, any randomly placed cell will include a random number $X_0$ of cluster 
centers.  That is, the distribution of $X_0$ will be Poisson, with a parameter 
that is specified by the epoch, labelled by $b$, at which the cell is placed, 
and by the size of the cell.  This model for the number of cluster centers in a 
randomly placed cell, given that the initial distribution was Poisson, is 
consistent with linear theory (e.g. Peebles 1980), and is motivated by a 
scaling argument that can be applied to the Poisson Press--Schechter 
description (section 3.2 in Sheth 1995).  However, the scaling argument is 
equivalent to the queueing theory interpretation of the Galton--Watson 
partition function of merger history trees (see discussion at the end of 
Appendix~C).  Thus, the assumption that the number of clusters in a randomly 
placed cell is a Poisson random variable is consistent with the partition 
structure derived earlier in this paper.  

When the distribution of $X_0$ is Poisson with parameter $\nbar_{\rm cl}$, then 
the probability that any randomly placed cell contains exactly $N$ particles is 
\begin{eqnarray}
f(N) &=& \sum_{m=0}^\infty p(m)\,P(N|m) = \sum_{m=0}^\infty
{\nbar_{\rm cl}^m {\rm e}^{-\nbar_{\rm cl}}\over m!}\,P(N|m) \nonumber \\
&=& {\nbar_{\rm cl}\over N!}\,\sum_{m=1}^\infty {N-1\choose m-1}\,
\nbar_{\rm cl}^{m-1}\,{(Nb)^{N-m}}\ {\rm e}^{-\nbar_{\rm cl}-Nb} \nonumber \\
&=& {\nbar_{\rm cl}\over N!}\,(\nbar_{\rm cl} + Nb)^{N-1}\,
{\rm e}^{-\nbar_{\rm cl} - Nb}.
\label{btfn}
\end{eqnarray}
Setting $\nbar_{\rm cl} = \bar nV(1-b)$, where $\bar n$ is the average density 
of particles and $V$ is the size of the cell, is required by normalization.  
This shows that equation~(\ref{btfn}) implies that the probability that a 
randomly placed cell of size $V$ contains exactly $N$ particles is
\begin{equation}
f(N,V) = {\nbar(1-b)\over N!}\ \Bigl[\nbar(1-b) + Nb \Bigr]^{N-1}\ 
{\rm e}^{-\nbar(1-b)-Nb} ,
\label{ppsd}
\end{equation}
where the left hand side now shows the volume dependence explicitly.

Equation~(\ref{ppsd}) is also a solution to the Saslaw \& Hamilton (1984) 
thermodynamic model of nonlinear gravitational clustering.  It can be 
understood as describing a Poisson distribution of point sized clumps, where 
the probability a clump has $N$ associated particles is given by a Borel 
distribution with parameter $b$ (Saslaw 1989).  Thus, equations~(\ref{borel}) 
and~(\ref{ppsd}) can be derived from a thermodynamic model 
(Saslaw \& Hamilton 1984; Sheth 1995), from an analysis of the excursion set 
statistics of overdense regions of a Poisson distribution (Sheth 1995), and 
from the Poisson Galton--Watson branching process (Consul 1989).  

In equation~(\ref{ppsd}), $b$ is constant, independent of cell size $V$.  This 
is a consequence of assuming that all clumps are point sized.  Relaxing this 
assumption means that the branching process is no longer straightforward to 
implement.  Nevertheless, we can use the Poisson cluster interpretation of 
this  branching process derivation of $f(N,V)$ to gain some understanding of 
the shape of the counts in cells distribution when the point sized 
approximation is relaxed.  The counts in cells distribution of a Poisson 
distribution of Borel (with parameter $b$) Press--Schechter clumps that have 
nontrivial sizes and shapes is well approximated by equation~(\ref{ppsd}), 
except that $b$ becomes scale dependent.  It tends to zero as $V\to 0$ and it 
tends to a constant value as $V$ becomes larger than the typical clump size 
(Sheth \& Saslaw 1994).  

Whereas the usual Press--Schechter analysis of excursion set mass functions  
provides information about the distribution of virialized clump masses, it does 
not provide information about the internal structure of these clumps, nor does 
it describe how these clumps are distributed relative to each other in space.  
The clumps may be correlated with each other, or distributed uniformly at 
random.  Thus, one cannot compute the $N$-point correlation functions of the 
clustered distribution, nor can one construct the nonlinear counts in cells 
distribution function.  In this respect, the Press--Schechter description does 
not provide a complete description of nonlinear clustering.  Therefore, it is 
very interesting that numerical simulations of gravitational clustering from an 
initially Poisson distribution confirm the accuracy of equation~(\ref{ppsd}) on 
all scales, as well as the scale dependence and temporal evolution of $b$ 
(Sheth \& Saslaw 1994 and references therein).  This measured accuracy of 
equation~(\ref{ppsd}), and the way in which it can be derived from the 
branching process, suggests one way in which the Press--Schechter approach may 
be extended to provide some information about the counts in cells distribution. 

Before concluding, we note that the partition structure derived in the main 
text is independent of the accuracy or applicability of the results of this 
Appendix.  That is, the results of the main text are independent of whether or 
not the Borel clusters have a Poisson spatial distribution.  

\end{document}